\documentclass{article}
\usepackage{amsmath,graphicx,url,amssymb,latexsym,epsfig,tabularx,setspace,verbatim,rotating,threeparttable}
\usepackage{natbib}
\usepackage[hmargin=1.5in,vmargin=1.5in]{geometry}

\begin{document}
\large
%\doublespace

\voffset 0.8in
\setcounter{secnumdepth}{4}
\begin{titlepage}
\centerline{\bf \LARGE Detection Technique for Artificially-Illuminated}
\centerline{\bf \LARGE Objects in the Outer Solar System and Beyond}

\vskip 0.7cm

\large
\centerline{\Large Abraham Loeb$^{1,2}$ and Edwin L. Turner$^{3,4}$}
\vspace{0.5cm}
\large

\noindent
$^1$ {\it Astronomy Department, Harvard University, 60 Garden Street,
Cambridge, MA 02138, USA}\\ 

\noindent
$^2$ {\it Institute for Theory and Computation, Harvard-Smithsonian Center
for Astrophysics, 60 Garden Street, Cambridge, MA 02138, USA}\\

\noindent
$^3$ {\it Department of Astrophysical Sciences, Princeton University,
Princeton, NJ 08544, USA}\\

\noindent
$^4$ {\it Institute for the Physics and Mathematics of the
Universe, The University of Tokyo, Kashiwa 227-8568, Japan}\\

%\vskip{1cm}
%\raggedright
%Corresponding author:\\
%Abraham Loeb\\
%Astronomy Department, Harvard University,
%60 Garden St., Cambridge, MA 02138, USA\\
%E-mail: aloeb@cfa.harvard.edu

\vskip 1cm

\large

\begin{abstract}
\large
 
Existing and planned optical telescopes and surveys can detect
artificially-illuminated objects comparable in total brightness to a
major terrestrial city out to the outskirts of the Solar System. Orbital
parameters of Kuiper belt objects (KBOs) are routinely measured to
exquisite precisions of $< 10^{-3}$.  Here we propose to measure the
variation of the observed flux $F$ from such objects as a function of
their changing orbital distances $D$.  Sunlight-illuminated objects
will show a logarithmic slope $\alpha \equiv (d\log F/d\log D)= -4$
whereas artificially-illuminated objects should exhibit $\alpha= -2$.
Planned surveys using the proposed LSST will provide superb
data that would allow measurement of $\alpha$ for thousands of KBOs.
If objects with $\alpha=-2$ are found, follow-up observations
can measure their spectra to determine if they are
illuminated by artificial lighting.  
%This method opens a new window in
%the search for extraterrestrial civilizations. 
The search can be
extended beyond the Solar System with future generations of telescopes on
the ground and in space, which would be capable of detecting phase
modulation due to very strong artificial illumination on the
night-side of planets as they orbit their parent stars.
\end{abstract}

\vskip 0.2in

\large
\noindent
{\bf Kewords:} astrobiology, SETI, Kuiper belt objects, artificial illumination

\end{titlepage}

\large
\section{Introduction}

The search for extraterrestrial intelligence (SETI) has been conducted
mainly in the radio band \citep{Wilson,Tarter,Shostak}, with
peripheral attention to exotic signals in the optical
\citep{Howard,Horowitz,Ribak,Dyson_Plant,Elvis} and thermal infrared
\citep{Dyson}.  Possible ``beacon'' signals broadcasted intentionally
by another civilization to announce its presence as well as the
''leakage'' of radiation, produced for communication or other purposes
({\it e.g.}, radar), have been the usual targets of radio SETI
observations.

As technology evolves on Earth, expectations for plausible
extraterrestrial signals change. For example, the radio power emission
of the Earth has been declining dramatically in recent decades due to
the use of cables, optical fibers and other advances in communication
technology, indicating that eavesdropping on distant advanced
civilizations might be more difficult than previously thought
\citep{Forgan}.

Here we are guided instead by the notion that biological creatures are
likely to take advantage of the natural illumination provided by the
star around which their home planet orbits.  As soon as such creatures
develop the necessary technology, it would be natural for them to
artificially illuminate the object they inhabit during its dark
diurnal phases.

Our civilization uses two basic classes of illumination: thermal
(incandescent light bulbs) and quantum (light emitting diodes [LEDs]
and fluorescent lamps).  Such artificial light sources have different
spectral properties than sunlight.  The spectra of artificial lights
on distant objects would likely distinguish them from natural
illumination sources, since such emission would be exceptionally rare
in the natural thermodynamic conditions present on the surface of
relatively cold objects.  Therefore, {\it artificial illumination may
serve as a lamppost which signals the existence of extraterrestrial
technologies and thus civilizations}.  Are there realistic techniques
to search for the leakage of artificial illumination in the optical
band?\footnote{Here we focus on illumination in the optical band but
identical considerations apply to creatures that evolved to sense
radiation in the UV and IR bands, in which stars are also highly
luminous.}

It is convenient to normalize any artificial illumination in flux
units of 1\% of the solar daylight illumination of Earth,
$f_\oplus\equiv 0.01(L_\odot/4\pi D_\oplus^2)= 1.4\times 10^4~{\rm
erg~s^{-1}~cm^{-2}}$, where $D_\oplus=1.5\times 10^{13}~{\rm cm}\equiv
1~{\rm AU}$ is the Earth-Sun distance. Crudely speaking, this unit
corresponds to the illumination in a brightly-lit office or to that
provided by the Sun just as it rises or sets in a clear sky on
Earth.\footnote{http://www.brillianz.co.uk/data/documents/Lumen.pdf}

\section{Artificially Illuminated Kuiper Belt Objects}

We first examine the feasibility of this new SETI technique within the
Solar System, which offers the best prospects for detecting
intrinsically faint sources of light.

The flux reaching an observer from any self-luminous source varies
according to the familiar inverse square law, but the flux from
scattered sunlight off an object at a distance $D\gg 1~{\rm AU}$
scales as $D^{-4}$ due to the combination of the inverse square
dependence of the solar flux which illuminates it combined with the
inverse square dependence of the scattered component of that incident
flux which reaches an observer on Earth.  Thus, the observed flux from
an object that is artificially illuminated at a level of $f_{\oplus}$
would be larger than the flux due to its reflected sunlight by a
factor of $(A/1\%)^{-1}(D/1~{\rm AU})^2$, where $A$ is the albedo
(reflection coefficient) of the object to sunlight.  The $A$ values of
objects in the outer solar system vary widely \citep{Albedos} and their
colors range from neutral to very red \citep{Colors}.  This implies
that the ratio of artificial illumination, with an unknown spectrum,
to scattered sunlight could be a strong function of wavelength.

More than $\sim 10^3$ small bodies have already been discovered in the
distance range of $30$--$50~{\rm AU}$, known as the Kuiper belt of the
Solar System \citep{Petit}. The number of known Kuiper belt objects
(KBOs) will increase by 1-2 orders of magnitude over the next decade
through wide-field surveys such as
Pan-STARRS\footnote{http://pan-starrs.ifa.hawaii.edu/public/home.html}
and LSST.\footnote{http://www.lsst.org/lsst/} The sizes\footnote{These
sizes correspond to diameters for the larger objects, which are
spherical in shape, but are merely characteristic linear scales for
the smaller objects which have irregular shapes.}  of known KBOs
($\sim 1$--$10^3~{\rm km}$) are usually inferred by assuming a typical
albedo \citep{Grundy} of $A\sim 4$--$10\%$. (The albedo of a KBO can
sometimes be calibrated more reliably based on measurements of its
thermal infrared
emission.\footnote{http://www.minorplanetcenter.org/iau/lists/Sizes.html})
For $A=7\%$ and a distance $D=50~{\rm AU}$, an artificially
$f_\oplus$-illuminated object would be brighter by a factor $\sim
3.6\times 10^2$ than if it were sunlight-illuminated. This implies
that an $f_\oplus$-illuminated surface would provide the same observed
flux $F$ as a sunlight-illuminated object at that distance, if it is
$\sim {\sqrt{3.6\times 10^2}}=19$ times smaller in size. In other
words, an $f_\oplus$-illuminated surface of size 53 km (comparable to
the scale of a major city) would appear as bright as a $10^3~{\rm km}$
object which reflects sunlight with $A=7\%$. Since $\sim 10^3~{\rm
km}$ objects were already found at distances beyond $\sim 50~{\rm
AU}$, {\it we conclude that existing telescopes and surveys could
detect the artificial light from a reasonably brightly illuminated
region, roughly the size of a terrestrial city, located on a KBO.}

Weaker artificial illumination by some factor $\epsilon<1$ relative to
the ``1\% of daylight on Earth'' standard represented by $f_\oplus$,
would lower the observed flux by the same factor, since the observed
flux scales as $F\propto \epsilon$.  Correspondingly, the equivalent
object size needed for artificial illumination to produce the same
observed flux as due to sunlight illumination, would increase by
$\epsilon^{-1/2}$. Nevertheless, {\it existing telescopes could
detect dimly illuminated regions ($\epsilon\sim 1\%$) hundreds of
km in size on the surface of large KBOs.}

The current artificial illumination on the night-side of the Earth has
an absolute $r$-band magnitude of roughly 44 (corresponding to
$1.7\times 10^{13}~{\rm lumens}$ produced from $\sim 2\times
10^{12}~{\rm Watts}$ of electric
power).\footnote{http://www.lightinglab.fi/IEAAnnex45/guidebook/11$\_$technical$\%$20potential.pdf}
\footnote{This value assumes a Sun-like spectrum in the optical band
and an illumination efficiency (lumens/watt) similar to that of the
Sun, which is in the range of modern fluorescent and LED lights as
well.  The choice of the $r$-band is obviously somewhat arbitrary and
is meant only for illustrative purposes.  The artificial illumination
employed by an alien civilization might have a wide range of possible
spectra, perhaps correlated with that of the primary star hosting the
object on which they evolved.}  Existing telescopes could see the
artificially-illuminated side of the Earth out to a distance of $\sim
10^3~{\rm AU}$, where its brightness in scattered sunlight and in
artificial lighting (at current levels) would coincidentally be
roughly equal.  A present-day major terrestrial city, Tokyo for
example,\footnote{http://www.tepco.co.jp/en/forecast/html/kaisetsu$-$e.html}
has an absolute $r$-band magnitude of very roughly 48 with apparent
$r$-magnitudes of approximately 16 at a distance of 1 AU, 24 at 30 AU,
26 at 100 AU and 31 (about as faint as the faintest detected objects
in the Hubble Ultra-Deep Field) at $10^3~{\rm AU}$.

Although precise numbers depend on many detailed properties of the
telescope, instrument, observing conditions (sky brightness, image
quality {\it etc.}), representative exposure times to reach the
aforementioned $r$-band apparent magnitudes at high (50-to-1)
signal-to-noise ratio are 1, 500 and 1800 seconds, respectively, for
the first three cases with an 8-meter class telescope in good
observing conditions and using modern CCD detectors.  Reaching $r\sim
31$ is not feasible from the ground and took over $3\times 10^5$
seconds with the 2.4-meter Hubble Space Telescope.

Thus, {\it existing optical astronomy facilities are capable of
detecting artificial illumination at the levels currently employed on
Earth for putative extraterrestrial constructs on the scale of a large
terrestrial city or greater out to the edge of the Solar System.}

\section{A Flux-Distance Signature of Artificial Illumination}

Orbital parameters of Kuiper belt objects (KBOs) are routinely
measured\footnote{Long-term monitoring of KBOs may also serve to
limit or detect deviations from Keplerian orbits due to artificial
propulsion.} to a precision of $< 10^{-3}$ via astrometric
observations \citep{Petit}. {\it A simple but powerful and robust
method for identifying artificially-illuminated objects is to measure
the variation of the observed flux $F$ as a function of its changing
distance $D$ along its orbit.}  Sunlight-illuminated objects will show
a logarithmic slope of $\alpha \equiv (d\log F/d\log D)= -4$ whereas
artificially-illuminated objects should exhibit $\alpha= -2$. The
required photometric precision of better than a percent for such
measurements (over timescales of years) can be easily achieved with
modern telescopes.

If objects with $\alpha=-2$ are discovered, follow-up observations
with long exposures on $8-10$ meter and space telescopes could
determine their spectra and test whether they are illuminated by
artificial thermal (incandescent) or quantum (LED/fluorescent) light
sources.\footnote{One should also examine images of the dark side of
Solar System moons, suspected of hosting liquid water.  For example,
city lights can be searched for in images taken by the Cassini
spacecraft of the dark side of Saturn's moon, Enceladus.}  The
exposure time requirements to achieve moderate signal-to-noise spectra
would be extreme, running to millions of seconds or more, at the faint
end of the magnitude range under consideration.  However, the
motivation to determine the nature and properties of an object showing
convincing $\alpha=-2$ behavior would be even more extreme.  A
complementary follow-up search for artificial radio signals could be
conducted with sensitive radio observatories \citep{Loeb}, such as
VLA,\footnote{http://www.vla.nrao.edu/}
ATA,\footnote{http://www.seti.org/ata}
GMRT,\footnote{http://gmrt.ncra.tifr.res.in/}
LOFAR,\footnote{http://www.lofar.org/}
MWA,\footnote{http://www.mwatelescope.org/} and
PAPER,\footnote{http://astro.berkeley.edu/$\sim$dbacker/eor/} which
would be able to detect extraordinarily low levels of radio emission
by current terrestrial standards.  In general, follow-up using all
available observational resources would be well justified.

KBOs vary in brightness for reasons other than their changing distance
from the Earth and the Sun \citep{Rab,Sch,Lightcurves}.  Specific
causes include a changing phase angle (due largely to the Earth's
orbital motion) leading to changes in the contributions from coherent
backscattering and surface shadowing, outgassing ({\it i.e.,} cometary
activity), rotation of objects with non-spherical shapes or surface
albedo variations, and for some objects occultation by a binary
companion.  Although the brightness changes associated with these
effects are typically tenths of a magnitude and can be larger for some
objects, their time scales are short (hours to days in most cases)
and, with the exception of outgassing, the resulting variations are
periodic.  For these reasons it will be necessary to monitor KBO
brightnesses frequently and for a period of years in order to model
or, at worst, average out other contributions to variability on an
object-by-object basis and allow the secular trend with changing
distance ({\it i.e.,} the $\alpha$ value) to emerge.  Fortunately,
LSST \citep{Ivezic} will obtain extensive and very high quality data
of precisely this nature for unrelated and conventional purposes.
Thus, {\it the survey we propose can identify KBO (or asteroid)
candidates for intensive follow-up with no investment of additional
observational resources.}

We note that artificial lights might also vary on short time scales,
either due to their being turned on and off, due to beaming, or due to
bright spots appearing and disappearing over the limb as the object
rotates.

\section{Night Lights Beyond the Solar System}

The next generation of ground-based telescopes
(EELT,\footnote{http://www.eso.org/public/teles-instr/e-elt.html}
GMT,\footnote{http://www.gmto.org/} and
TMT\footnote{http://www.tmt.org/}) as well as space telescopes
(JWST,\footnote{http://www.jwst.nasa.gov/}
Darwin,\footnote{http://www.esa.int/export/esaSC/120382$\_$index$\_$0$\_$m.html}
and
TPF\footnote{http://planetquest.jpl.nasa.gov/TPF/tpf$\_$index.cfm})
will be able \citep{Ria} to search for artificial illumination of
extra-solar planets \citep{Sch1, Sch2}. Although the $\alpha$ test
proposed above for objects in the outer Solar System is not relevant
for exoplanets, a search for the orbital phase (time) modulation of
the observed flux from the artificial illumination of the night-side
on Earth-like planets as they orbit their primary could be used in its
place. The observer would see stronger artificial illumination when
the dark side of the planet is more in view, exactly the opposite of
the case with natural day side illumination from the star.  Cloud
cover would mask some of the artificial illumination of an
Earth-like planet in a stochastic time dependent manner, which might
significantly complicate the interpretation of such phase curves.

A preliminary broad-band photometric detection could be improved
through the use of narrow-band filters which are tuned to the spectral
features of artificial light sources (such as LEDs).  For this
signature to be detectable, the night side needs to have an artificial
brightness comparable to the natural illumination of the day side.
Clearly, the corresponding extraterrestrial civilization would need to
employ much brighter and more extensive artificial lighting than we do
currently since the global contrast between the day and night sides is
a factor $\sim 6\times 10^5$ for the present-day Earth.  In favorable
scenarios, some proposed versions of NASA's TPF mission would have
reasonable prospects of detecting the artificial illumination of an
exoplanet if it were at levels a few times greater than $f_\oplus$ or
more.

City lights would be easier to detect on a planet which was left in
the dark of a formerly-habitable zone after its host star turned into
a faint white dwarf. The related civilization would need to survive the
intermediate red giant phase of its star. If it does, separating its
artificial light from the natural light of a white dwarf, would be
much easier than for the original star, both in contrast and in
absolute brightness.

\section{Concluding Remarks}

In addition to the low prior probability that should probably be
assigned to the idea of an alien civilization occupying KBOs, the
search proposed in this paper could fail for a host of other plausible
reasons.  The artificially illuminated spaces might be underground or
otherwise shielded for a variety of reasons, such as to avoid wasting
of energy or to maintain a stealthy presence.  Advanced technology,
including biological alteration of sensory organs, might be employed
to render very low natural illumination levels useable.  Moreover, the
most easily detectable signatures might well be in very different
bands, such as radio emissions.  Thus, as for all other known SETI
techniques, a null result would have no clear meaning.  However, this
is not a sufficient reason to refrain from searching since it is
clearly impossible to predict the behaviors or capabilities of unknown
alien civilizations with any confidence and because a positive
result would carry such immense implications.

Artificially-lit KBOs might have originated from civilizations near
other stars. In particular, some small bodies may have traveled to the
Kuiper belt through interstellar space after being ejected dynamically
from other planetary systems \citep{Moro}. These objects can be
recognized by their hyperbolic orbits.  A more hypothetical origin for
artificially-lit KBOs involves objects composed of rock and water/ice
(asteroids or low-mass planets) that were originally in the habitable
zone of the Sun, developed intelligent life, and were later ejected
through gravitational scattering with other planets (such as the Earth
or Jupiter) into highly eccentric orbits. Such orbits spend most of
their time at their farthest (turnaround) distance, $D_{\rm max}$. If
this distance is in the Kuiper belt, then the last time these objects
came close to Earth was more than $\sim 500~(D_{\rm max}/10^2~{\rm
AU})^{3/2}$ years ago, before the modern age of science and
technology began on Earth.

\vskip 0.45in
\noindent
{\bf ACKNOWLEDGEMENTS.} We thank F. Dyson, M. Holman and A. Parker for
helpful comments.  AL was supported in part by NSF grant AST-0907890
and NASA grants NNX08AL43G and NNA09DB30A.  ELT gratefully
acknowledges support from a Princeton University Global Collaborative
Research Fund grant and the World Premier International Research
Center Initiative (WPI Initiative), MEXT, Japan.

\newpage

\bibliographystyle{apj}

\end{document}